\documentclass[a4paper]{jpconf}
\usepackage{graphicx}
\usepackage{lineno}
\usepackage[T1]{fontenc}
\begin{document}
\title{Delayed charge recovery discrimination of passivated surface alpha events in P-type point-contact detectors}

\author{J.\,Gruszko$^1$, on behalf of the \textsc{Majorana} Collaboration}

\address{$^{1}$Center for Experimental Nuclear Physics and Astrophysics, and Department of Physics, University of Washington, Seattle, WA, USA}

\ead{jgruszko@uw.edu}

\begin{abstract}
The \textsc{Majorana Demonstrator} searches for neutrinoless double-beta decay of $^{76}$Ge using arrays of high-purity germanium detectors. If observed, this process would demonstrate that lepton number is not a conserved quantity in nature, with implications for grand-unification and for explaining the predominance of matter over antimatter in the universe. A problematic background in such large granular detector arrays is posed by alpha particles. In the \textsc{Majorana Demonstrator}, events have been observed that are consistent with energy-degraded alphas originating on this surface, leading to a potential background contribution in the region-of-interest for neutrinoless double-beta decay. However, it is also observed that when energy deposition occurs very close to the passivated surface, charges drift through the bulk onto that surface, and then drift along it with greatly reduced mobility. This leads to both a reduced prompt signal and a measurable change in slope of the tail of a recorded pulse. In this contribution we discuss the characteristics of these events and the development of a filter that can identify the occurrence of this delayed charge recovery, allowing for the efficient rejection of passivated surface alpha events in analysis.
\end{abstract}

\section{Introduction}
The \textsc{Majorana Demonstrator} \cite{Elliott} \cite{MJD_AHEP} is an ultra-low-background array of high-purity Germanium (HPGe) detectors, which searches for the neutrinoless double-beta decay ($0\nu\beta\beta$) of $^{76}$Ge. The experiment is composed of 44.8\,kg of PPC detectors, 29.7\,kg of which is enriched to 88\% $^{76}$Ge. The first of its two modules began data-taking in 2015, with the second beginning commissioning in the summer of 2016. The primary goal of the \textsc{Demonstrator} is to demonstrate a path forward to achieving a background rate at or below 1\,count/(ROI\,t\,y) (in the 4\,keV region of interest (ROI) around the $0\nu\beta\beta$ Q-value of 2039\,keV) in a next-generation Ge-based experiment. 

\section{Alpha Backgrounds in the \textsc{Majorana Demonstrator}}
\subsection{Surface Alpha Mitigation}
To achieve the \textsc{Demonstrator}'s background goals, backgrounds from $\alpha$ decays near or on the detectors must be strictly controlled. $\alpha$ decays that originate from radon progenies are particularly problematic, as radon may plate onto detector surfaces during manufacture and assembly. Due to implantation in the surrounding materials or incomplete charge collection, the resulting high-energy alpha interactions can lead to a significant background in the $0\nu\beta\beta$ ROI.  To reduce radon implantation on the detector and mount surfaces, the detectors were assembled and installed into cryostats inside a glovebox with a radon-reduced dry N$_{2}$ environment, and all mounting parts were acid-etched or leached and placed into equivalent storage. 

The geometry of PPC detectors provides further mitigation of surface alpha events relative to, for example, bolometric arrays. The majority of the detector surface is composed of a ~1\,mm thick Li$^{-}$-diffused n$^{+}$ contact, which $\alpha$ particles cannot penetrate. The only detector surfaces that are sensitive to $\alpha$ backgrounds are those of the small (approx. 3\,mm diameter) p$^{+}$ contact and the passivated surface of the crystal. The charge collection properties near this surface can differ for different detector models. 

\subsection{Delayed Charge Recovery (DCR) in Surface Alpha Events}
\begin{figure}[htb]
\begin{minipage}{.7\textwidth}
\includegraphics[width=\linewidth]{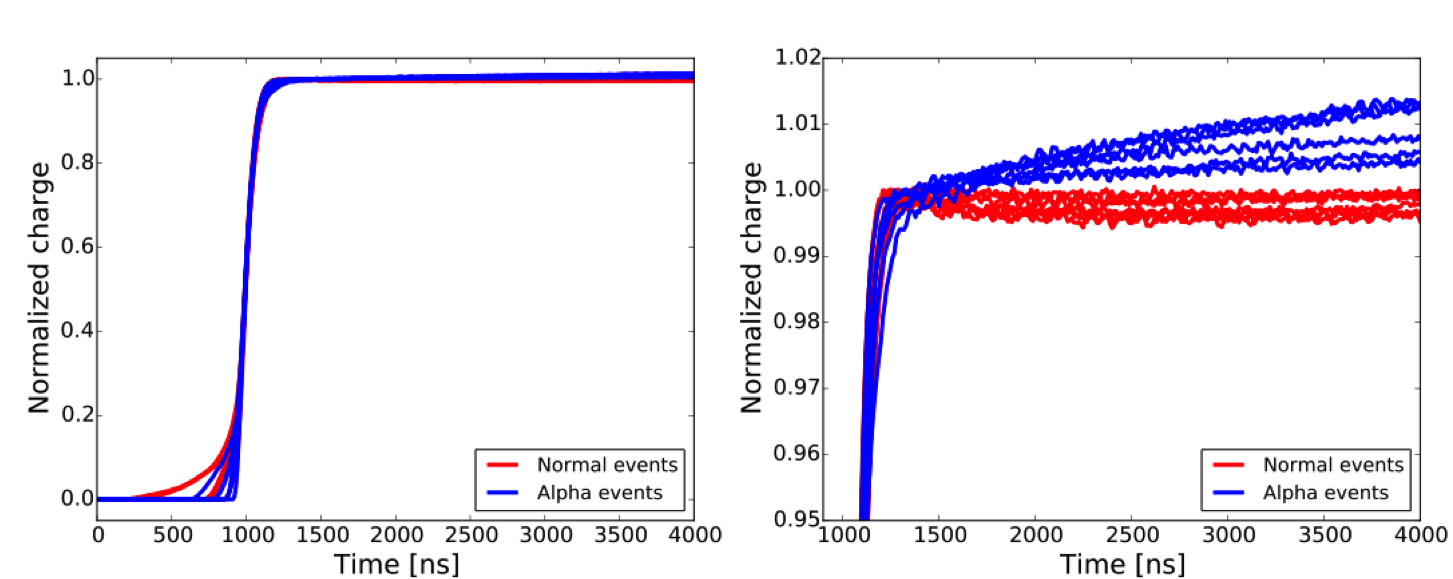}
\end{minipage}\hspace{2pc}%
\begin{minipage}{.20\textwidth}
\caption{\label{alpha_wf}Sample pole-zero corrected waveforms. $\alpha$ event waveforms occurring near the passivated surface, in blue, show a distinctive slow component.}
\end{minipage}
\end{figure}

In the PPC detector geometry, the electric field in the region of the passivated surface may lead to charge drift along the surface in addition to the normal bulk transport of charge. As seen in Monte Carlo simulations of hole drift in HPGe crystals \cite{Mullowney}, surface charges drift 10 to 100 times more slowly than bulk charges. This leads to the addition of a slow component to the usual fast (bulk) pulse. The energy of these events is degraded, since a significant fraction of the charge is not collected promptly. The presence of the slow component also leads to a measurable change in the slope of the signal tail, as seen in Fig.\,\ref{alpha_wf}. 

\
Detectable $\alpha$ interactions are most likely to occur at the passivated surface, since the range of $\alpha$s with typical energies is less than $20~\mu m$ in germanium \cite{ASTAR}. Therefore these events are far more likely to demonstrate DCR effects than $\gamma$ or $\beta$ interactions. Tagging and rejecting these events removes a large fraction of $\alpha$ background with minimal sacrifice of bulk events.    
\subsection{DCR Parameter Implementation}
\begin{figure}[h]
\begin{minipage}{.28\textwidth}
\includegraphics[width=\linewidth]{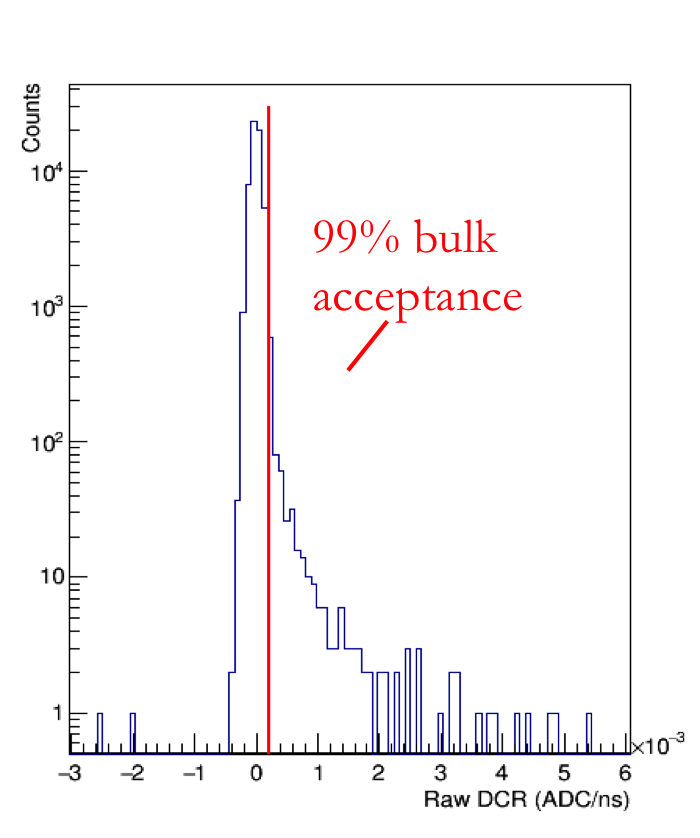}
\caption{\label{rawDCR}Raw DCR for single-site calibration events with energy above 1\,MeV in one detector. Events falling to the right of the line are rejected.}
\end{minipage}\hspace{1pc}%
\begin{minipage}{.4\textwidth}
\includegraphics[width=\linewidth]{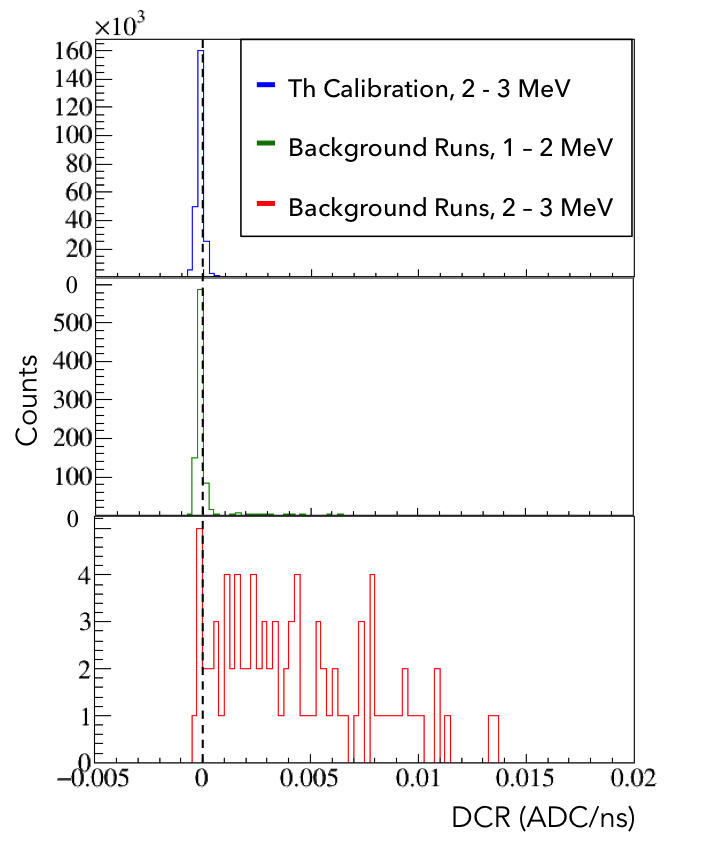}
\end{minipage}
\begin{minipage}{.28\textwidth}
\caption{\label{DS1} 90\% of $\gamma$ and electron events fall to the left of the dashed line and $\alpha$ events preferentially fall to the right. During a calibration run {\it(top)}, $\gamma$ events survive the cut. In background runs, events in the energy range dominated by $\beta\beta(2\nu)$ survive the cut and candidate $\alpha$ events {\it(bottom)} are removed. } 
\end{minipage}
\end{figure}

The DCR parameter is calculated using $^{228}$Th calibration Compton continuum events with $E>1$\,MeV. Single-site events are selected using A/E \cite{Cuesta} and pile-up cuts. To first order, the exponential decay of normal bulk events leads to a tail slope of $\Delta=\frac{aE}{\tau}$ where $\tau$ is the decay constant of the tail, $E$ is the energy of the pulse, and $a$ is a scaling constant. Using a linear fit to $\Delta$ vs. $E$, $\Delta$ is projected onto the energy axis to give the raw DCR parameter: DCR$_{raw}$$\,=\Delta-(\frac{aE}{\tau}+b)$. The bulk-acceptance efficiency of the cut is set as desired, and the DCR parameter is shifted such that events with DCR$_{corr}<0$ are accepted by the cut. See Fig.\,\ref{rawDCR}.

The bulk acceptance of the cut was set at $90\pm3.7\%$ to optimize the $0\nu\beta\beta$ sensitivity. The uncertainty of the efficiency is evaluated by quantifying two main systematic effects: the energy dependence and the pulse-shape dependence. The first of these was determined by summing, in quadrature, the residuals of the efficiency in 40\,keV bins. The second was taken to be the absolute difference between bulk acceptance in the $^{208}$Tl double-escape peak and the nominal acceptance. The statistical uncertainty is negligible, and the systematic effects are assumed to be independent.

\subsection{Data Set 1 Results}
In the \textsc{Majorana Demonstrator} Data Set 1 (see \cite{Elliott}), which consists of 662.2\,kg-days of exposure (607.4\,kg-days of enriched exposure), the DCR cut retains 83.6\% of events with energy between 1 and 2\,MeV, an energy region dominated by $2\nu\beta\beta$ events. It removes 88.4\% of events with energy between 2 and 3\,MeV. The DCR parameter shows clear separation between $\gamma$/$\beta$ events and $\alpha$ events, as seen in Fig.\,\ref{DS1}. 

\section{Conclusion}
Based on preliminary results, the DCR filter appears to effectively remove $\alpha$ background events. Coming improvements, including the use of a longer digitization window and the addition of a fast rise time cut that will identify p$^{+}$ events, should further improve $\alpha$ rejection. Collimated $\alpha$ source measurements are underway to fully characterize the \textsc{Majorana} PPC detector response to surface $\alpha$ events.

\ack
This material is based upon work supported by the NSF Graduate Research Fellowship Program under Grant No. DGE 1256082. This material is based upon work supported by the U.S. Department of Energy, Office of Science, Office of Nuclear Physics. We acknowledge support from the Particle Astrophysics and Nuclear Physics Programs of the National Science Foundation. This research uses these US DOE Office of Science User Facilities: the National Energy Research Scientific Computing Center and the Oak Ridge Leadership Computing Facility. We acknowledge support from the Russian Foundation for Basic Research. We thank our hosts and colleagues at the Sanford Underground Research Facility for their support.


\section*{References}
\begin{thebibliography}{5}
\bibitem{Elliott} Elliott S {\it et al.} 2016 {\it J. Phys.:\,Conf. Ser.} {\bf these proceedings}
\bibitem{MJD_AHEP} Abgrall N {\it et al.} 2014 {\it Adv. High Energy Phys.} {\bf 2014} 365432
\bibitem{Mullowney} Mullowney P {\it et al.} 2012 {\it Nucl. Instr. Meth. Phys. Res.} {\bf A662} 33--44
\bibitem{ASTAR} Berger, M\,J, {\it et al.} 2005 {\it ASTAR} [Online] Available: http://physics.nist.gov/Star
\bibitem{Cuesta} Cuesta C {\it et al.} 2016 Heavy Quarks and Leptons 2016, Blacksburg VA {\it Preprint} {\tt arXiv:1608.07340}
\end{thebibliography}
\end{document}